\documentclass[]{aa}
\usepackage{graphicx}

\begin{document}

\thesaurus{01(08.05.3; 08.16.3; 10.05.1; 10.07.2}

\title{Theoretical vs.\ semi-empirical relative ages of globular clusters}

\author{L.~Pulone\inst{1,2} \and M.~Salaris\inst{3,4} \and A.~Weiss\inst{3} \and
R.~Buonanno\inst{1}}

\institute{Osservatorio Astronomico di Roma, Via dell'Osservatorio 2,
00040 Monte Porzio Catone (Roma), Italy
\and European Southern Observatory, Karl-Schwarzschild-Str.~2, 85748
Garching, Germany 
\and Max-Planck-Institut f\"ur Astrophysik,
Karl-Schwarzschild-Str.~1, 85748 Garching, Germany
\and
Astrophysics Research Institute, Liverpool John Moores University,
Byrom Street, Liverpool L3 3AF, UK}  
\offprints{L.~Pulone}
\mail{L.Pulone}
\date{Received ; accepted}

\maketitle

\begin{abstract} 
Theoretical relative ages of galactic globular clusters have recently
been challenged by a semi-empirical relation. It was used to point
out that tested sets of iso\-chrones were unable to reproduce the
relation and yield internally inconsistent relative ages.
We find that differential cluster ages derived with 
the isochrones by Salaris \&
Weiss (1998) are reliable and internally consistent. 
We also show that this consistency depends on using the lower absolute 
ages determined by SW98, 
which therefore receive more empirical support.
Moreover, we discuss the effect of the clusters absolute age 
on the evaluation of their differential ages, and its
connection with the question of their age dispersion. 

\end{abstract}

\section{Two opponent views of relative cluster ages}

The determination of galactic globular cluster (GC) ages has recently
come into a state of flux, first due to updated stellar physics and lately
due to the HIPPARCOS results. Theoretical isochrones were
challenged by  Buonanno et al.\ (1998, hereafter BCP), who have presented a
self-consistent method for the determination of relative cluster ages,
which as much as possible makes use of observational properties and
tries to minimize the input from theoretical isochrones. They argue that
three tested sets of isochrones fail to reproduce the empirical
relation between $\triangle (B-V)_{\rm TO}^{\rm RGB}$ (the (B-V) difference between 
the TO and the base of the RGB)
and [Fe/H] at a
given age. Furthermore, relative ages based on brightness or colour
differences are inconsistent in all cases. In this
{\em Letter} 
we will confront the isochrones by Salaris \& Weiss (1997, 1998;
hereafter SW97 \& SW98) with the observational results and show that
this set passes the test to a great extent.

SW97 have determined GC ages by means of the traditional theoretical
approach of comparing theoretical iso\-chrones with an observed CMD.
The differences with respect 
to related works are (i) that they use the very latest
(canonical) input physics, (ii) that also $\alpha$-element enhancement
in the GC composition is taken into account (e.g.\ in the
opacity tables) and (iii) that they determined {\em
absolute} ages only for the very few clusters where both CMD
morphology and available data guarantee an accurate dating. 
The brightness difference between Turn-Off (TO) and Zero-Age
Horizontal Branch (ZAHB), $\triangle V^{\rm HB}_{\rm TO}$, was used as
the absolute age indicator. The imposed
constraints result in a rather small sample of  7 clusters of
all metallicities (including disk GC in SW98)
suitable for the determination of absolute ages (see SW97 and SW98
for details). All other ages in their sample of 31 GC have been
determined differentially with respect to these template clusters by 
using the (theoretical)
dependence of $\triangle (B-V)_{\rm TO}^{\rm RGB}$ 
on age.
For minimizing the errors arising from uncertainties in the adopted
colour transformations and the mixing length calibration, SW97 and
SW98 evaluated the relative ages
only within metallicity groups. Whether the relative
ages would also be valid across 
metallicity group boundaries, SW97 checked in only one
case. This and other checks (e.g.\ availability of two clusters with
an absolute age determination in the same metallicity group, 
comparison of the template clusters distances
obtained from the theoretical ZAHB models with main-sequence fitting 
distances using HIPPARCOS subdwarfs) resulted in the confirmation
of the internal consistency of the SW97 and SW98 ages and of the reliability 
of their theoretical isochrones. 

We now recall briefly the BCP method for determining homogeneous
relative ages for GC. The first step is to define a sample of coeval
clusters in a wide range of metallicities. This requires age determinations, which
at least must be able to define correctly what is ``coeval''. To this scope,
BCP used a variant of the vertical method used by SW97, employing a point
on the main-sequence 0.05 mag redder than the TO. The sensitivity of
this new absolute age indicator ($\triangle V^{0.05}$) and its
reliability have been investigated in great detail in BCP. In total, a
sample of 11 clusters covering ${\rm [Fe/H] } = -2.2 \cdots -0.5$ was
defined, whose absolute ages agreed within $\pm$1 Gyr. While the
absolute age of this group depended on the set of isochrones used
(D'Antona et al.\ 1997; VandenBerg 1996, private communication to BCP;
Straniero \& Chieffi 1996, p.c.\ to BCP),
membership within the group was basically invariant. From this, the empirical
relation of $\triangle (B-V)_{\rm TO}^{\rm RGB}$ vs.\ $\rm [Fe/H]$ (at
the common age of the group of clusters) is obtained (Fig.~11 of BCP),
which in turn can be compared to the predicted one. Here, BCP noted
that none of the adopted isochrone sets agreed with the empirical
relation. It was also demonstrated that the discrepancy is independent of
the colour-transformations employed and thus the inconsistency could
arise from the theoretical models themselves.

In the second step 6 clusters of definitely lower age than that of the
calibrating group have been used to determine the change in $\triangle
(B-V)_{\rm TO}^{\rm RGB}$ with age at given $\rm[Fe/H]$. The
ages of the young clusters relative to the calibrating ones were
obtained by means of the $\triangle V^{0.05}$ method. Again,
theoretical models have to be used for this step; only theoretical
predictions for the stellar brightness, but not for colours, are used. 
This finally yielded
an empirical expression relating age differences with two observational 
quantities, namely $\triangle (B-V)_{\rm TO}^{\rm RGB}$ and metallicity
(Eq.~3 and Fig.~16 of BCP).
This relationship is then applied for deriving relative ages for GC 
whose ZAHB level cannot be properly defined (blue or scarcely populated HB).
For all other clusters in the BCP sample 
the average of the $\triangle V^{0.05}$ and the $\triangle (B-V)_{\rm
TO}^{\rm RGB}$ relative ages is used.

Considering that the BCP relationship depends itself on theoretical isochrones  
which were demonstrated to be inconsistent in at least one respect,
it is worthwhile to question the reliability of the BCP-relation, which
can be valid only if the theoretical isochrones do not suffer from
systematic errors in the definition of the coeval clusters.
Second, since SW97 and SW98 claim the internal consistency of their
results, both in 
terms of absolute and relative ages, one could ask
whether this result depends on the smaller sample of clusters selected
by SW97 and SW98.  
In the following we will therefore re-derive the BCP relationship
using the SW98 isochrones where theoretical input is necessary. 

\begin{figure}[ht]
\includegraphics[scale=0.53,draft=false]{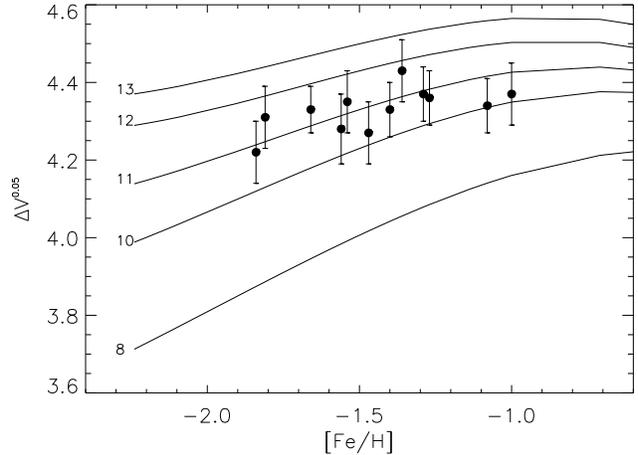}
\caption{$\triangle V^{0.05}$ for the 12  coeval clusters
compared to the theoretical values by SW98 for several ages
(in Gyr)} 
\label{f:1}
\end{figure}

\section{Data set and coeval sample}

The reference GC data set is that of BCP with the addition of NGC~6171
(a metal-rich halo GC included in SW97) and
the exclusion of the three disk GC 47~Tuc, M~71 and NGC~6352. 
As extensively discussed in Alonso et al.\ (1997), 
$\triangle (B-V)_{\rm TO}^{\rm RGB}$ and $\triangle V^{\rm HB}_{\rm
TO}$ yield significantly different ages for these clusters, if the
typical GC helium content 
($0.24$) is assumed. The problem can be resolved with a nearly solar
value ($0.27$; SW98) for the disk clusters,
but this would add an unwanted additional degree
of freedom for the semi-empirical relationship investigated here.
Moreover, given the 
strong dependence of the horizontal-branch luminosity
on the He abundance, the inclusion of these clusters 
would affect the present comparison 
with the models because it would affect the selection of the sample of 
coeval clusters. The reader should be well aware of this difference
between BCP and the present paper,  
because at the metal-rich end of the relationship 
$\triangle (B-V)_{\rm TO}^{\rm RGB} vs {\rm [Fe/H]}$
BCP found 
the largest discrepancies between observations and models.
	
Adopting this data set we have selected a sample of coeval clusters, on the 
base of the observed $\triangle V^{0.05}$ values and their errors, and
the absolute ages as determined from the SW98 isochrones. 
The largest possible sample of 
coeval clusters is different from BCP and it
is shown in Fig.~\ref{f:1} (see also data in 
Tab.~\ref{t:1});
individual ages range from 10 to 12 Gyr and
overlap within the errors, providing an average age of 10.9 Gyr.
The sample spans the metallicity range -2.0$<[Fe/H]<$-1.0, 
typical of the bulk of halo GC, and, contrarily to BCP,  
it does not include any of the most metal poor 
clusters with [Fe/H]$<$-2.0. This is in agreement with the results by SW97 and SW98 
who found that the most metal poor GC are on average older than the other GC. 

The result that the coeval clusters are not the same as in BCP is
at odds with their conclusion that the membership to the coeval group does not 
depend on the isochrones considered. 
The main reason for this occurrence is the lower 
absolute age found for the individual GC with the SW98 isochrones. As previously described, 
the average age  of the coeval GC is 10.9 Gyr, while BCP considered an average age 
of 15 Gyr.
As it is evident 
from Figs. 7 and 8 in BCP, the shape of the theoretical relation between 
$\triangle V^{0.05}$ and [Fe/H] depends on the absolute age of the isochrones.
For example, at ages around 12 Gyr, the derivative $\delta \triangle
V^{0.05}/\delta({\rm age})$
at low metallicities is quite different 
from the case of 14 or 16 Gyr, independently of the set of colour transformations 
and bolometric corrections used.

If we would have shifted by $\approx$ +0.2 mag the $\triangle V^{0.05}$ values
of the isochrones used in BCP (see their Fig. 8) in such a way that the average age of 
the coeval clusters is around 11 Gyr, we would have excluded the most metal poor 
ones from the coeval sample, thus recovering the result obtained with the SW98 isochrones.

\begin{table}
\caption{Data for the coeval GC sample}
\begin{flushleft}
\begin{tabular}{llll}
\noalign{\smallskip}
\hline
\noalign{\smallskip}
  $Cluster$  &  $\triangle V^{0.05}$ &$\triangle (B-V)_{\rm TO}^{\rm RGB}$ & [Fe/H]\\
\noalign{\smallskip}
\hline
\noalign{\smallskip}
NGC~362        & 4.36$\pm$0.07 & 0.263$\pm$0.01& -1.27\\
NGC~1261       & 4.37$\pm$0.07 & 0.261$\pm$0.01& -1.29\\
NGC~3201       & 4.28$\pm$0.09 & 0.243$\pm$0.01& -1.56\\
NGC~5272 (M~3) & 4.33$\pm$0.06 & 0.249$\pm$0.01& -1.66\\
NGC~5904 (M~5) & 4.33$\pm$0.07 & 0.250$\pm$0.01& -1.40\\
NGC~6101       & 4.31$\pm$0.08 & 0.239$\pm$0.01& -1.81\\
NGC~6121 (M~4) & 4.43$\pm$0.08 & 0.248$\pm$0.01& -1.36\\
NGC~6171       & 4.37$\pm$0.08 & 0.260$\pm$0.01& -1.00\\
NGC~6362       & 4.34$\pm$0.07 & 0.264$\pm$0.01& -1.08\\
NGC~6584       & 4.35$\pm$0.08 & 0.248$\pm$0.01& -1.54\\
Pal~5          & 4.27$\pm$0.08 & 0.266$\pm$0.01& -1.47\\
Arp~2          & 4.22$\pm$0.08 & 0.248$\pm$0.01& -1.84\\
\noalign{\smallskip}
\hline 
\label{t:1}
\end{tabular}
\end{flushleft}
\end{table}

Another effect of the lower absolute age of the GC is that now
we find included in the coeval sample two clusters, Pal~5 and Arp~2, 
considered as young by BCP.

\section{Differential ages}  

The first step to evaluate GC relative ages following the technique by BCP
is the calibration of a relation
between $\triangle (B-V)_{\rm TO}^{\rm RGB}$ and metallicity
for the subset of coeval clusters. 
A linear fit to the data (taking into account the
$\triangle (B-V)_{\rm TO}^{\rm RGB}$ error given in Tab.~\ref{t:1} and a
typical error of $\pm$0.15 dex in metallicity) provides

\begin{equation}
\triangle (B-V)_{\rm TO}^{\rm RGB} = (0.028\pm0.013){\rm [Fe/H]}
+(0.294\pm0.020)
\end{equation}

The next step is the evaluation of the ratio $\delta / \Delta t_9$, where
$\delta$ is the difference between the observed $\triangle (B-V)_{\rm
TO}^{\rm RGB}$ and the one {\em expected} on the basis of Eq.~(1), and
$\Delta t_9$ is the relative age (in Gyr) with respect to the coeval clusters,
as determined from $\triangle V^{0.05}$.
We determined this ratio
for the four younger clusters NGC~1851,  Pal~12, Rup~106 and
Ter~7, (see BCP for a discussion), then averaged 
these four values (having checked that they are not significantly correlated with
metallicity) and get:
\begin{equation}
\delta / \Delta t_9=-0.0158\pm0.0057 \rm{~mag/Gyr}
\end{equation}
This quantity is smaller than that of BCP (-0.0093), implying a larger
sensitivity of $\triangle (B-V)_{\rm TO}^{\rm RGB}$ to age and
therefore smaller age differences for an observed $\triangle
(B-V)_{\rm TO}^{\rm RGB}$.
By combining Eq.~(1) and  Eq.~(2) we obtain the final relation between
$\triangle (B-V)_{\rm TO}^{\rm RGB}$, [Fe/H] and $\Delta t_9$:
\begin{eqnarray*}       
\Delta t_9&=& (-0.0158\pm 0.0057)^{-1} \cdot [(\triangle (B-V)_{\rm TO}^{\rm RGB}\\
&-& (0.028\pm0.013) {\rm [Fe/H]} - (0.294\pm 0.020)]  ~~~~~~~(3a) 
\end{eqnarray*}
This equation allows the proper calculation of the error in $\Delta t_9$ 
by standard error propagation. 
Taking into account the errors in $\triangle (B-V)_{\rm TO}^{\rm RGB}$
and [Fe/H] as well, one gets errors of the order of  
$\pm$2.0 Gyr in the derived values of $\Delta t_9$ (see the vertical error bars
in Fig.~\ref{f:3}).
Without the errors Eq.~(3a) can be simplified to
\begin{eqnarray*}
\Delta t_9 = -63.3 \triangle (B-V)_{\rm TO}^{\rm RGB} + 1.8 {\rm [Fe/H]}
+ 18.6  ~~~~~~~~~~(3b)
\end{eqnarray*}
The coefficients depend of course
on the absolute age of the
coeval sample; for the sake of comparison we recall here that
the corresponding values obtained by BCP were (-107.5, 4.3, 33.3).

\section {Discussion}

\begin{figure}
\includegraphics[scale=0.45,draft=false]{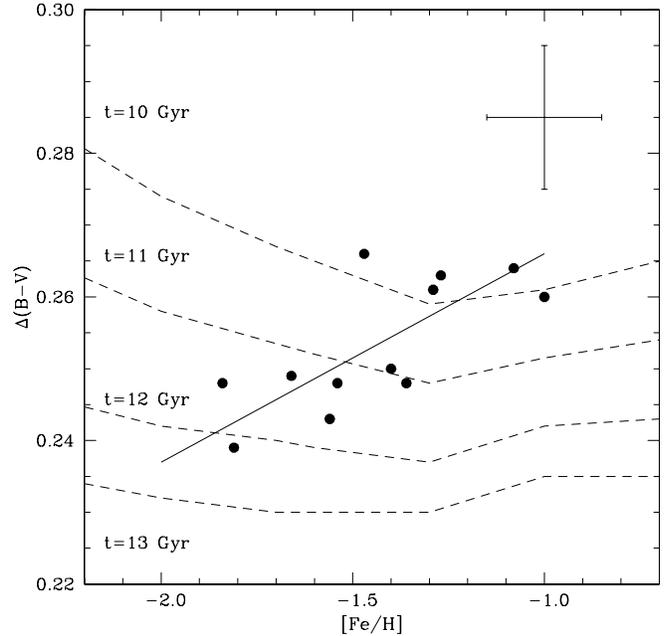}
\caption{Predictions of the SW98 isochrones (dashed lines ) for
$\triangle (B-V)_{\rm TO}^{\rm RGB}$ as a function of [Fe/H], compared
with Eq.~(1) (solid). Filled circles correspond
to the observational points for the coeval clusters. A typical error
bar is shown as well)}
\label{f:2}
\end{figure}

In Sect.~1 we recalled that BCP found an inconsistency in the
theoretical isochrones by comparing the empirical relation Eq.~(1)
with them. In Fig.~\ref{f:2} we show the same comparison and,
in the considered range of metallicity, we did not find the
same problem: the empirical relation crosses a 2-Gyr range in the
isochrones, between 10 and 12 Gyr, which is identical to the variation
of ages determined from $\triangle V^{0.05}$ (Fig.~\ref{f:1}).
It is also evident that the spread of the observational points and the errors 
associated to them (see Table~1) is large enough to make the 11 Gyr isochrone
fully compatible with the empirical relation. 

\begin{figure}
\includegraphics[scale=0.43,draft=false]{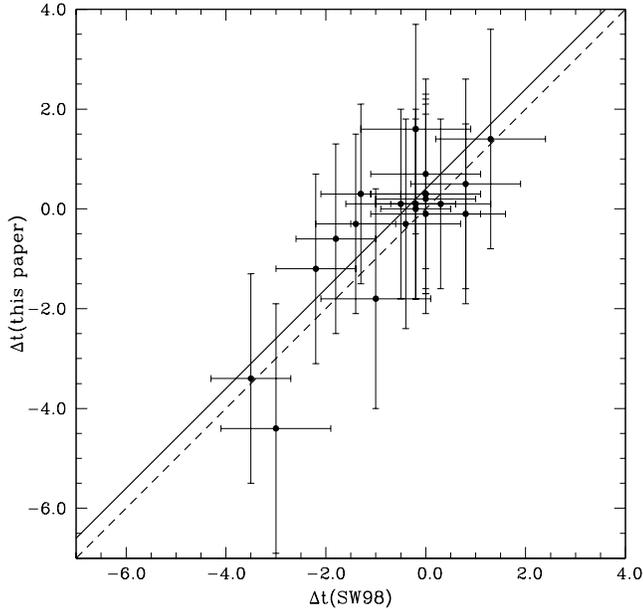}
\caption{Comparison of the GC differential ages (in Gyr; with respect to
t=10.9) as obtained by SW98 and from
Eq.~(3a). The dashed line displays the 1:1 relation, the solid one the
regression line of the points.}
\label{f:3}
\end{figure}

To further substantiate our claim that the SW98 iso\-chrones -- and therefore	
the semi-empirical relation derived in the previous section -- 
are trustworthy, we show in	
Fig.~\ref{f:3} the comparison between the differential ages 	
(with respect to t=10.9 Gyr) obtained with both
methods for the 21 clusters in common (the coeval GC with the exception of 
NGC6121 and NGC6362, the 4 young ones and the blue HB clusters 
NGC288, NGC1904, NGC2298, NGC6254, NGC6397, NGC6752 and NGC7492)
with $-2.0< {\rm [Fe/H]} <-1.0$ on the 
Zinn \& West (1984) metallicity scale, as adopted in the data set. 
Before doing so, we had to correct the ages 
found by SW98 for the different metallicity scale they have used, namely 
the Carretta \& Gratton (1997) one. This 
new metallicity scale lowers the absolute ages by $\approx$0.8 Gyr 
when used instead of the
Zinn \& West (1984) one (SW98). Therefore, we have increased by this amount
the ages given in SW98, and we have computed their differential ages 
with respect to the zero-point of the semi-empirical 
relative ages, fixed at 10.9 Gyr.

The agreement between the two sets of differential ages is
excellent: there is just a zero point shift  
of $0.4 \pm 0.5$  Gyr, well within the errors associated with the age 
determinations. This small zero point shift reflects basically the fact that 
for the coeval clusters in common (10 out of 12), 
SW98 find an average age (after the correction for the 
different metallicity scale) of 10.6 Gyr instead of 10.9 Gyr. 
When considering only the 7 blue HB clusters, one finds a zero point
shift of $0.7\pm0.8$ Gyr, in good agreement with the results from 
the complete sample of 21 clusters. 

The dependence of Eqs.~(3a-b) on the absolute age of the GC
affects also the derived age spread
among the clusters. To quantify whether a real intrinsic age
dispersion exists among the clusters, over that due to the
errors in the single age determinations, we have performed
the F-test (Press et al.\ 1992) in the
way described by Chaboyer et al.\ (1996) and SW97.  It compares the
{\em observed} age distribution with the {\em expected} 
one under the assumption of no intrinsic age dispersion,
and provides an estimate of
the probability that the two distributions are actually the same,
taking into account
the errors on the individual ages. If
this test indicates that the sample is not coeval, the size of the
true age dispersion $(\sigma_{\rm real})$ among the clusters can be
estimated according to $\sigma_{\rm real}^2={(\sigma^{2}_{\rm
obs}-\sigma^{2}_{\rm exp})}$, where $\sigma_{\rm obs}$ and $\sigma_{\rm exp}$
are, respectively, the 1$\sigma$ dispersion in the observed age distribution and in the 
expected one.
When considering the calibration of the semi-empirical method given by our
Eqs.~(3a-b), we find that the complete sample of GC 
(with the exclusion of the clusters with
[Fe/H]$<$-2.0, for which we could not calibrate the method) shows 
a real 1$\sigma$ age spread by 0.9 Gyr.
Performing the same test with the calibration of the method given in BCP
for the same sample of GC, one obtains a 1$\sigma$ 
spread by $\approx$ 2 Gyr.

To conclude, we find that both the theoretical and the semi-empirical
approach yield the same relative ages for the cluster sample of
BCP when using the isochrones by SW98. In particular, the 
$\triangle (B-V)_{\rm TO}^{\rm RGB}$ values as a function of age and metallicity
predicted by these models
are fully consistent with the observations.
We also showed how the empirical method and results deduced from it (e.g.\
the amount of the age spread among GC) depend on the absolute
age of the coeval sample and pointed out that the definition
of this sample also depends on the GC absolute ages.
The consistency of relative ages derived from the isochrones
(Figs.~\ref{f:1} and \ref{f:2}) and the agreement with those from the
empirical relation is achieved only for the lower ages determined in
SW97 and SW98. This result therefore further substantiates the lower ages for GC
as found recently in several papers.

\begin{acknowledgements}
The work of M.S. was carried out as part of the TMR
programme 
financed by the EC.
\end{acknowledgements}


\begin{thebibliography}{}
\bibitem[10]{asm.97} Alonso A., Salaris M., Mart\'{\i}nez-Roger C.,
Straniero O., Arribas S., 1997, A\&A 323, 374
\bibitem[11]{bcp98} Buonanno R., Corsi C.E., Pulone L., Fusi Pecci F.,
Bellazzini M., 1998, A\&A 333, 505, BCP
\bibitem[15]{cg97} Carretta E., Gratton R.G. 1997, A\&AS 121, 95
\bibitem[16]{cds96} Chaboyer B., Demarque P., Sarajedini A., 1996, ApJ 459, 558
\bibitem[14]{dm97} D'Antona F., Caloi V., Mazzitelli I., 1997, ApJ
447, 519
\bibitem[17]{nrec} Press W.H., Teukolsky S.A., Vetterling W.T., Flannery
B.P. 1992, Numerical Recipies (Cambridge Univ. Press)
\bibitem[12]{sw97} Salaris M., Weiss, A., 1997, A\&A 327, 107, SW97
\bibitem[13]{sw98} Salaris M., Weiss, A., 1998, A\&A 335, 943, SW98
\bibitem[18] {zw84} Zinn R., West M.J., 1984, ApJS 55, 45 
\end{thebibliography}
\end{document}